# Characterizing the Failure Modes of LLMs in Resolving Real-World GitHub Issues

Yanjie Jiang, Yian Huang, Guancheng Wang, *Member, IEEE,* Junjie Chen, Hui Liu, Lionel Briand, *Fellow, IEEE*

***Abstract*—Large Language Models (LLMs) are increasingly deployed to resolve real-world GitHub issues. However, despite their potential, the specific failure modes of these models in complex repair tasks remain poorly understood. To characterize how LLM behavior diverges from human developer practices, this paper evaluates three state-of-the-art models, i.e., *Claude 4.5 Sonnet*, *Gemini 3 Pro*, and *GPT-5*, on the *SWE-bench Verified* dataset. We conduct a rigorous manual analysis of the symptoms and root causes underlying 243 failed attempts across 900 total trials. Our investigation first yields a unified failure taxonomy encompassing five distinct stages of the repair pipeline, within which we categorize typical failure symptoms and their prevalence. Secondly, our findings reveal that for all evaluated LLMs, *strategy formulation and logic synthesis* constitutes the most error-prone stage, followed by *problem understanding*, whereas *localization* exhibits the lowest failure rate. This suggests that LLMs may excel at fault localization, a task traditionally regarded as one of the most formidable challenges in automated program repair. Furthermore, we observe that robustness and operational costs (particularly in failure scenarios) vary significantly across different models. Finally, we uncover the root causes of these failures and propose actionable strategies to mitigate them. A particularly notable finding is that existing evaluation harnesses occasionally misjudge correct patches due to superficial discrepancies or hidden constraints. Collectively, our insights may provide promising directions for enhancing the effectiveness and reliability of LLM-based issue resolution.**

## I. Introduction

THE evolution of Large Language Models has catalyzed a fundamental paradigm shift in automated program repair by transitioning from the generation of isolated code snippets to the autonomous resolution of repository-level software issues [1]–[3]. Contemporary agents powered by frontier LLMs are now expected to navigate large-scale codebases, manage cross-module dependencies, and reason over long-range execution contexts [4], [5]. These capabilities extend beyond local code synthesis toward full-fledged software engineering tasks that demand both advanced reasoning and system-level understanding.

Despite rapid progress, current LLM-based repair systems remain far from human-level reliability. Benchmarks such as SWE-bench [1] provide evaluation settings grounded in real-world issues and pull requests. However, recent work [6] suggests that performance on such benchmarks may be inflated due to limitations in test suites. Even so, according to the latest leaderboard data as of April 30, 2026 [7], state-of-the-art models resolve only around 76.8% of verified issues. This persistent gap highlights a substantial disparity between model performance and the proficiency of expert developers.

More importantly, current evaluation protocols provide limited visibility into the underlying causes of these failures [8]–[10]. Existing studies primarily focus on improving benchmark pass rates through stronger prompting, more sophisticated agents, or enhanced tool usage. At the same time, the evaluation itself remains dominated by a binary pass/fail metric based on final test execution [11]–[14]. This reductionist view simplifies a complex, multi-stage reasoning process into a single aggregate score, limiting insights from autonomous agent research.

As a result, autonomous repair research faces a fundamental interpretability gap. While we can measure whether a model succeeds, we cannot explain how its reasoning diverges from human debugging behavior. However, repository-level issue resolution is inherently a multi-stage, trajectory-driven task that requires coordinated reasoning across problem understanding, localization, strategy formulation, and implementation [10], [15], [16]. Consequently, existing studies lack a systematic, process-level understanding of how and where LLM-driven repair deviates from human debugging behavior. Moreover, prior work rarely grounds failure analysis in direct comparisons with human-written patches, making it difficult to determine whether observed errors reflect general debugging challenges or fundamentally distinct machine-specific failure modes.

This gap motivates a more fundamental question: `How do LLMs fail differently from human developers?` A failed resolution attempt may not reflect an inability to generate code, but rather a mismatch between machine reasoning and human software engineering expertise. Human developers rely on architectural intuition, implicit framework conventions, system-wide invariants, and domain-specific judgment when debugging real-world software. In contrast, LLM agents operate through iterative prompting, localized observations, and test-driven feedback [11], [12], [17]. Without systematically com-

Yanjie Jiang, Yian Huang, and Junjie Chen are with School of Computer Software, Tianjin University, Tianjin 300350, China (e-mail: yanjiejiang@tju.edu.cn; 3023244073@tju.edu.cn; junjiechen@tju. edu.cn;) *Corresponding author: Junjie Chen*

Guancheng Wang is with the Research Ireland Lero Centre for Software, University of Limerick, V94 T9PX Limerick, Ireland (e-mail: guancheng.wang@ul.ie);

Hui Liu is with the School of Computer Science and Technology, Beijing Institute of Technology, Beijing 100081 (e-mail: liuhui08@bit.edu.cn);

Lionel Briand is with the University of Ottawa, Ottawa, ON K1N 6N5, Canada, and also with the Research Ireland Lero Centre for Software, University of Limerick, V94 T9PX Limerick, Ireland (e-mail: lbriand@uottawa.ca, lionel.briand@lero.ie).

paring AI-generated patches against reference patches written by human developers, it remains unclear whether LLMs merely replicate common debugging mistakes or introduce fundamentally different cognitive failure modes.

To address this challenge, we conduct an empirical study of failure patterns in autonomous repository-level issue resolution. Using the SWE-bench Verified dataset, we evaluate three frontier models (Claude 4.5 Sonnet, Gemini 3 Pro, and GPT 5) under a standardized interactive framework based on mini-SWE-agent [18], [19]. We deliberately adopt this execution environment to isolate intrinsic model reasoning from complex agentic scaffolding. We perform three independent trials for 100 tasks, resulting in 900 repair attempts, and conduct an in-depth manual root-cause analysis of 243 failed cases by examining execution trajectories, generated patches, reference patches, and evaluation traces.

To systematically investigate the resolution gap between autonomous agents and human developers, we formulate the following research questions:

**RQ1: How are failures distributed across the agentic issue resolution pipeline, and to what extent do LLMs' root causes align with or diverge from human error patterns?** By establishing a unified taxonomy and comparing it with human developer behavior, we can identify whether LLMs introduce novel failure modes or mirror human cognitive biases.

**RQ2: What behavioral patterns do LLMs exhibit during failures, and how do execution stability and trajectory efficiency reflect their reasoning capabilities?** Beyond outcomes, analyzing variance across runs and redundant reasoning loops provides insight into robustness and practical deployment costs.

**RQ3: What task-specific features characterize persistent failures, and what targeted interventions can help bridge the expertise gap between agents and human experts?** By linking failure modes to intrinsic task characteristics, we can propose actionable strategies to enhance the robustness and reliability of the next generation of issue resolution.

To answer these questions, we perform a fine-grained analysis of repair trajectories by jointly examining model actions, generated patches, and human reference solutions. This enables us to move beyond aggregate benchmark outcomes and identify where repair processes break down, providing a principled basis for understanding model limitations and for improving future issue-resolution systems.

Our analysis reveals several findings that differ from the bottlenecks commonly reported in earlier coding paradigms. First, pure fault localization and mechanical tooling errors are relatively rare. Instead, most failures are cognitive in nature. Models frequently struggle with strategy formulation, often producing partial fixes or destructive architectural rewrites, and they exhibit strong alignment bias by anchoring on speculative user hints. Second, we identify a limitation in benchmark evaluation itself: strict evaluation criteria and rigid test harnesses may incorrectly reject semantically correct patches, conflating syntactic matching with functional equivalence. Third, while frontier models share similar cognitive and evaluation vulnerabilities, they exhibit distinct execution stability characteristics and consistently suffer from context inefficiencies during failed reasoning trajectories. Finally, by correlating these failure modes with task-specific features, we demonstrate that explicitly exposing hidden evaluation constraints and boundary conditions through targeted prompt interventions can mitigate evaluation-related failures, whereas overcoming domain knowledge gaps requires architectural enhancements, such as explicit retrieval of external documentation.

In summary, this paper makes the following contributions:

- **A unified failure taxonomy for autonomous agents.** We construct a classification framework tailored to repository-level repair, covering five stages: problem understanding, localization, strategy & logic, implementation & execution, and validation & harness constraints.
- **Characterization of divergent failure patterns.** We systematically identify where LLM behavior diverges from human reasoning. For example, models excel at mechanical localization, while they struggle with strategic rule synthesis, which contradicts many prevailing assumptions in automated program issue resolution.
- **Actionable insights for agent architectures.** We derive practical implications for future agent designs by showing that proving models with missing contextual information and hidden task constraints is often more critical for success than simply increasing raw model intelligence.
- **Replication package.** We release a comprehensive dataset including 900 execution traces and 243 expert-annotated failure reports to support future investigations into model reasoning behaviors (Replication Package will release after publication).

The remainder of this paper is organized as follows: Section II reviews related work. Section III details the study design and Section IV presents the empirical results. Section V synthesizes the findings into broader implications and discusses threats to validity. Finally, Section VI summarizes the conclusion.

## II. Related Work

### A. Automated Issue Resolution

The integration of large language models has transformed automated coding tasks. Early neural approaches primarily framed code modification as a machine translation task. They used models such as CodeT5 [20] and PLBART [21] to map faulty or incomplete code snippets directly to their target versions. As these models scale in capability, conversational and pipeline methods emerged. These allowed models to iteratively generate and refine code by utilizing compiler feedback and test execution results [17], [22].

However, resolving real-world GitHub issues involves much more than fault localization. It requires addressing complex tasks at the repository level, including bug fixes, feature additions, and architectural modifications. To tackle this, recent research has shifted to autonomous-agent frameworks. Systems such as SWE-agent [11], AutoCodeRover [12], and Agentless [13] equip language models with specific search and navigation tools. This enables them to autonomously explore codebases, retrieve relevant context, and apply edits across multiple files. A critical prerequisite for these complex workflows is translating a natural-language issue description into a test case or script that reproduces the issue. Early approaches, such as LIBRO [23], use static prompt engineering to generate candidate tests, while more recent frameworks, such as AEGIS [24] and EvoCoder [25], employ interactive agent loops and execution feedback to dynamically refine testing code. Tools like AssertFlip [26] further optimize this process by generating passing tests first and mutating them to capture the target behavior.

As these multi-step frameworks grow in complexity, recent empirical studies have begun to formalize their workflows to better understand their internal mechanics and output quality. For example, Ceka et al. [9] investigate the decision-making pathways of software agents through the lens of traceability, conducting a direct comparison between agent-generated patches and human developer solutions to identify alignment gaps. Notably, their findings highlight the need to extend fault taxonomies to cover a broader range of agent configurations. Our study fills this gap by systematically classifying failures across the complete issue resolution pipeline. Concurrently, Bouzenia et al. [10] shift the analytical focus toward the execution process itself. By structuring agent interactions into thought-action-result trajectories, they provide a methodological framework for analyzing reasoning coherence and step-by-step decision-making, which informs our trajectory-based diagnostic approach.

While these autonomous frameworks and subsequent empirical analyses significantly advance the field of automated issue resolution, their heavy scaffolding often intertwines external engineering heuristics with the language model's native capabilities. This tight coupling obscures the intrinsic limitations of the underlying models, making it unclear whether failures stem from tooling deficits or fundamental reasoning bottlenecks. Therefore, our study intentionally strips away complex search heuristics. By utilizing mini-SWE-agent, a minimalist execution environment that removes components such as planning and iterative repair mechanisms [19], we isolate and assess the intrinsic reasoning limitations of frontier models during repository-level patch generation.

### B. Limitations of Existing Failure Analyses

Understanding why automated coding techniques fail is essential for advancing the field. For example, some researchers have investigated the failure of program repair techniques and identified operational bottlenecks related to search space explosion and imprecise fault localization [27]. They also successfully identified the test suite overfitting phenomenon associated with program repair [14], in which program repair techniques generate plausible but semantically incorrect heuristic patches merely to bypass crashing tests.

As language models have emerged as the primary engine for software engineering tasks, researchers have begun to investigate their failure characteristics. However, existing empirical studies predominantly focus on static models performing isolated code generation or localized bug fixing. For instance, researchers have observed that models struggle with basic context limits or produce localized semantic mistakes when evaluated on function-level datasets [15], [28]–[30]. Similarly, studies such as that of Liu et al. [8] highlight reasoning gaps but primarily view them through the lens of traditional patch-generation workflows. However, such analyses exhibit a critical limitation. They treat the language model as a simple text generator rather than an autonomous agent interacting with a complex environment. Autonomous agents operating in full repository environments face fundamentally different challenges. They need to translate ambiguous natural language into actionable strategies, navigate directories, and manipulate existing architectures. Existing taxonomies fail to capture the complex cognitive breakdowns and strategic missteps that occur when agents attempt to resolve real-world issues.

### C. Benchmarks for Code Evaluation

The evolution of autonomous coding agents is closely tied to the complexity of their evaluation environments. Early code generation was primarily evaluated on function-level datasets like HumanEval [31] and MBPP [32]. Similarly, traditional repair techniques relied on datasets such as Defects4J [33] and QuixBugs [34], which isolated the evaluation to predefined files and deterministic unit tests.

To bridge the gap between academic benchmarks and real-world software development, the community has introduced repository-scale benchmarks such as RepoBench [35], DevBench [16], and the SWE-bench series [1]. SWE-bench transforms the evaluation paradigm by requiring models to resolve general GitHub issues directly from natural language reports within full repository environments. These tasks span both bug fixes and feature enhancements.

However, designing an empirical study to analyze model reasoning requires a dataset in which failures can be confidently attributed to the model rather than to flawed data. Recent critiques of the original SWE-bench highlight that automated benchmark construction introduces substantial noise [36]. Many issues in the original dataset lack sufficient context to be solved. Others contain brittle test harnesses, environmental pollution, and overly rigid test oracles that enforce arbitrary historical implementation details rather than functional correctness.

To ensure the validity of our failure mode analysis, we therefore select the SWE-bench Verified dataset [37] as our experimental testbed. This dataset is a vetted subset of the original benchmark, where human software engineers manually annotate and confirm that each issue contains sufficient context for resolution and that it is evaluated using a fair, deterministic evaluation harness [38]. By utilizing this verified dataset, we systematically control for benchmark noise. This ensures that the failures observed in our study accurately reflect the models' limitations across diverse issue-resolution tasks.

## III. Study Design

### A. Overview of the Empirical Workflow

To systematically investigate our research questions, we formulate an end-to-end empirical workflow, as illustrated in Figure 1. The methodology is structured into three phases:

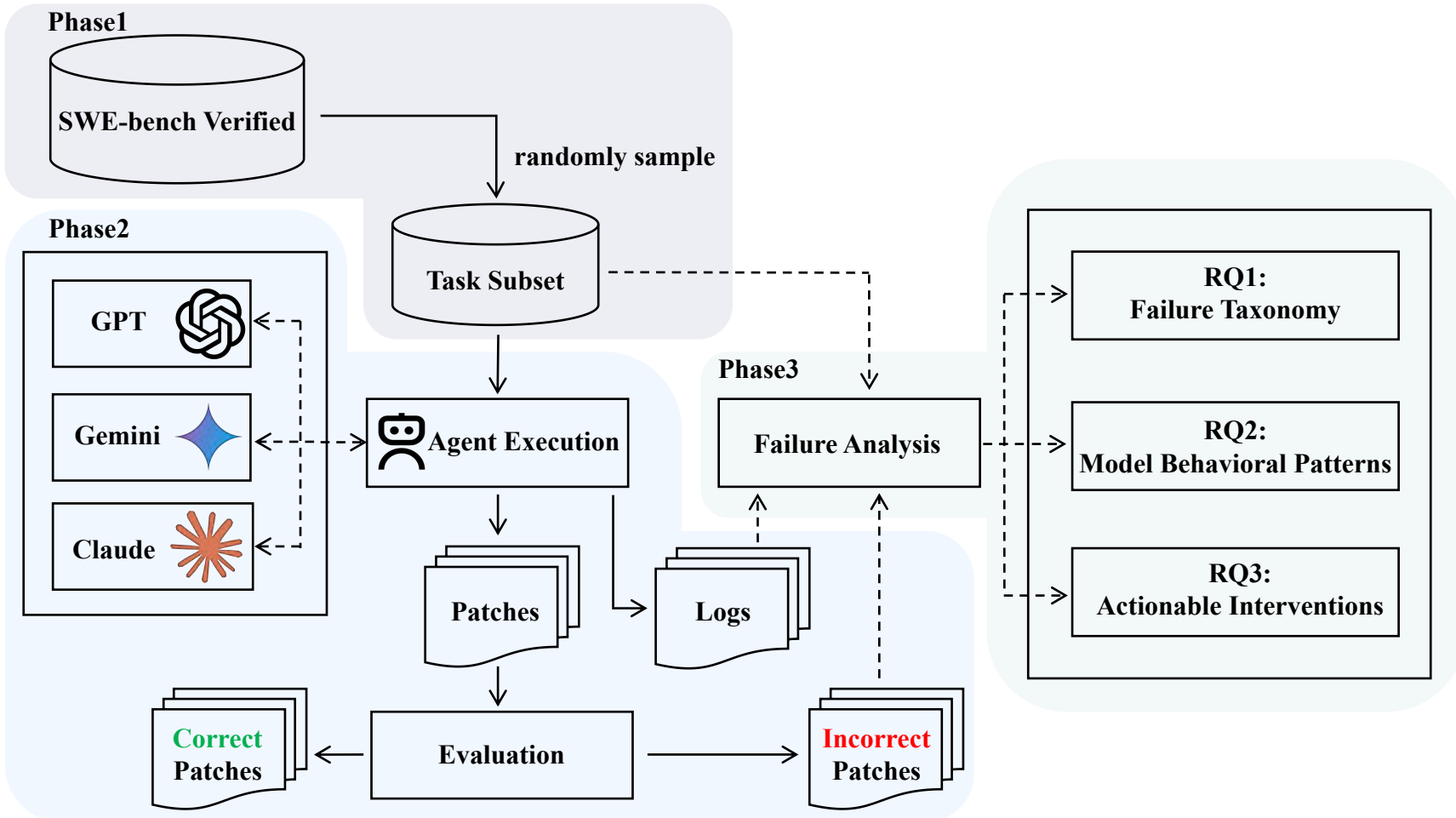

Fig. 1. Workflow of Our Empirical Study.

- **Phase 1: Data Preparation.** The process initiates with the construction of our experimental corpus. We perform random sampling from the SWE-bench Verified dataset [37] to ensure a representative task subset (Section III-B).
- **Phase 2: Autonomous Execution and Evaluation.** This corpus is processed by the agent execution framework (Section III-C), which orchestrates three state-of-the-art frontier LLMs (Section III-D) to generate both candidate patches and comprehensive interaction logs. Each patch undergoes rigorous validation via the benchmark's test harness, where results are distinguished into successful resolutions or persistent failures (Section III-E).
- **Phase 3: Multi-dimensional Failure Analysis.** This is the core of our methodology (Section III-F). We synthesize evidence from three distinct layers: task specifications, agent-environment interaction logs, and the semantic patterns of erroneous patches. By triangulating these sources, we formalize a comprehensive failure taxonomy (RQ1), characterize model-specific behavioral patterns (RQ2), and distill actionable interventions to bridge the expertise gap between autonomous agents and human experts (RQ3).

### B. Experimental Dataset and Task Selection

For this empirical study, we employ the SWE-bench Verified dataset [37], [39]. A rigorous root-cause analysis requires a granular qualitative inspection of agent trajectories, code patches, and execution traces. To balance the depth of human-in-the-loop inspection with the practical constraints of manual analysis, we select a representative subset of *100 tasks* from the original *500-task* benchmark for in-depth analysis.

To ensure the sample is unbiased and reproducible, we apply a deterministic random selection to the official test split. Specifically, we apply a deterministic shuffle (`seed=42`) to the official test split using the built-in `-shuffle` functionality of `mini-SWE-agent`, subsequently selecting the first 100 tasks via the `-slice` parameter. This procedure ensures that the sampling process remains transparent and can be exactly replicated by future researchers. The statistical consistency between our experimental subset and the entire dataset is summarized in Tables I and II.

TABLE I
ESTIMATED FIX TIME DISTRIBUTION.

| Estimated Fix Time | #Sample | Sample % | #Official | Official % | Δ (pp) |
|---|---|---|---|---|---|
| <15 min | 30 | 30.0 | 194 | 38.8 | -8.8 |
| 15 min–1 hour | 61 | 61.0 | 261 | 52.2 | +8.8 |
| 1–4 hours | 9 | 9.0 | 42 | 8.4 | +0.6 |
| >4 hours | 0 | 0.0 | 3 | 0.6 | -0.6 |

TABLE II
REPOSITORY DISTRIBUTION.

| Repository | #Sample | Sample % | #Official | Official % | Δ (pp) |
|---|---|---|---|---|---|
| django__django | 50 | 50.0 | 231 | 46.2 | +3.8 |
| sympy__sympy | 8 | 8.0 | 75 | 15.0 | -7.0 |
| sphinx-doc__sphinx | 10 | 10.0 | 44 | 8.8 | +1.2 |
| matplotlib__matplotlib | 9 | 9.0 | 34 | 6.8 | +2.2 |
| scikit-learn__scikit-learn | 7 | 7.0 | 32 | 6.4 | +0.6 |
| astropy__astropy | 4 | 4.0 | 22 | 4.4 | -0.4 |
| pydata__xarray | 3 | 3.0 | 22 | 4.4 | -1.4 |
| pytest-dev__pytest | 4 | 4.0 | 19 | 3.8 | +0.2 |
| pylint-dev__pylint | 3 | 3.0 | 10 | 2.0 | +1.0 |
| psf__requests | 1 | 1.0 | 8 | 1.6 | -0.6 |
| mwaskom__seaborn | 0 | 0.0 | 2 | 0.4 | -0.4 |
| pallets__flask | 1 | 1.0 | 1 | 0.2 | +0.8 |

To characterize task complexity, we utilize the *estimated fix time* metadata provided by the benchmark [37], [39]. This metric reflects the time required by an experienced software engineer to implement a resolving patch after codebase familiarization. As detailed in Table II, we track the repository-wise distribution using absolute counts (#Sample) and proportions (Sample%), comparing them against the original counts (#Official, Official%). The deviation Δ *(pp)* demonstrates that our subset preserves the structural characteristics of the complete benchmark, thereby supporting the generalizability of our findings. The complete list of `instance_id` for these 100 tasks is included in our replication package (Replication package will release after publication) to facilitate future verification.

### C. Execution Framework

To isolate the intrinsic reasoning and coding capabilities of LLMs from the influence of complex agentic scaffolding, we adopt `mini-SWE-agent` [18] as our execution framework. This choice is also consistent with the official SWE-bench Verified leaderboard [7], enabling direct comparison with state-of-the-art baselines.

Compared with more sophisticated agent frameworks, mini-SWE-agent provides a simpler and more transparent execution environment. It relies on standard *bash* commands rather than proprietary tool interfaces, reducing unnecessary abstraction and making model behavior easier to interpret. In addition, each action is executed independently using `subprocess.run`, improving execution stability and enhancing reproducibility across repeated runs. More importantly, the framework records a complete trajectory log for every execution, including prompts, shell commands, model responses, and environment feedback in a fully sequential manner. These fine-grained execution traces are essential for post-hoc failure analysis, as they allow us to systematically inspect the repair process and accurately identify the underlying causes of model failures.

### D. Models and Execution Settings

*1) Models and Inference Settings:* We evaluate three frontier large language models using an identical agent interface:

- `Claude: claude-sonnet-4-5-20250929` [40]
- `Gemini: gemini-3-pro-preview` [41]
- `GPT: gpt-5-2025-08-07` [42]

We select these models because they represent the top-performing proprietary models within their respective families on the SWE-bench Verified leaderboard as of December 1st, 2025. This selection serves two purposes. First, it provides strong and practically relevant representatives of the major frontier model ecosystems currently used in real-world software engineering workflows. Second, to control for differences in capability across model families, we select the strongest available model from each family as its representative. Rather than comparing a state-of-the-art model from one family with a substantially weaker model from another, this design enables a more meaningful comparison of failure behaviors during the repair process. We use the exact model versions listed on the leaderboard to maximize reproducibility. All decoding parameters follow the default settings of the corresponding model APIs. All experiments are conducted using `mini-SWE-agent (V1.15.0)`.

*2) Execution Budgets and Repeated Trials:* Within the mini-SWE-agent framework, we begin with its default execution configuration, which limits each attempt to a maximum of 250 interaction steps and a wall-clock timeout of 300 seconds for each executed shell command [18].

During pilot experiments, we observe that the default per-attempt cost limit of $3.00 occasionally causes complex tasks to terminate before the agent produces a final patch submission. To avoid premature interruption while preserving a reasonable safety bound, we increase the cost limit to $5.00. This threshold is used solely as a safeguard against abnormal execution behavior, such as infinite loops, rather than as an optimization parameter. Notably, across all 900 attempts in our study, no execution is forcefully terminated by the step limit, timeout threshold, or cost budget. We verify this by inspecting the complete execution logs and confirming that every run ends with an explicit model-generated submission decision rather than external truncation. This ensures that the observed failures reflect model limitations rather than artificial budget constraints.

To address the inherent non-determinism in multi-step agent trajectories even under fixed decoding parameters, we perform `three independent trials` for every task across all models. This results in a total of 900 attempts (100 tasks $\times$ 3 models $\times$ 3 trials). This multi-trial design enables us to distinguish between incidental failures caused by stochastic branching and systematic limitations that reflect a fundamental expertise gap in autonomous software repair.

### E. Evaluation Procedure and Success Criteria

We evaluate all generated patches using the official SWE-bench evaluation service [43], which provides a standardized and deterministic execution environment based on fixed repository snapshots and containerized test infrastructure. This ensures that all issue-resolution attempts are assessed under identical conditions and that evaluation outcomes are fully reproducible across repeated runs. During the repair process, the language model produces a code modification, referred to as the generated patch. For evaluation, we submit this patch together with its corresponding task identifier (`instance_id`) to the official evaluation service. The service then applies both the generated patch and the developer-provided test patch (`test_patch`) to the historical repository snapshot and executes the relevant test suite within the standardized environment.

Following the official SWE-bench protocol, a repair attempt is considered successful only if it passes all tests defined in the evaluation harness. Specifically, the generated patch should both resolve the target defect, meaning that all `FAIL_TO_PASS` tests pass, and avoid introducing regressions, meaning that all `PASS_TO_PASS` tests continue to pass. Notably, all non-perfect outcomes, including partial bug fixes, regression-inducing patches, and attempts that fail to produce a valid final patch, are treated uniformly as unsuccessful resolutions in the evaluation. However, in our failure analysis, these failed attempts are further examined and categorized by their specific failure modes. Importantly, this strict evaluation criterion serves not only as a performance metric but also as the foundation for comprehensive failure analysis. By treating even a single test failure as an unsuccessful attempt, we can capture the full spectrum of model breakdowns rather than just obvious repair failures. Combined with the complete execution trajectories recorded by the agent framework, this enables us to align failing outcomes with specific reasoning steps and systematically identify the underlying causes of failure.

### F. Failure Diagnosis Workflow

To diagnose failure modes, we systematically analyze artifacts produced throughout the issue-resolution pipeline, in-cluding the issue description, the generated patch, the developers' reference patch, evaluation traces, and complete agent interaction logs. Based on these sources, we establish a standardized manual analysis procedure to ensure consistent and reproducible failure attribution.

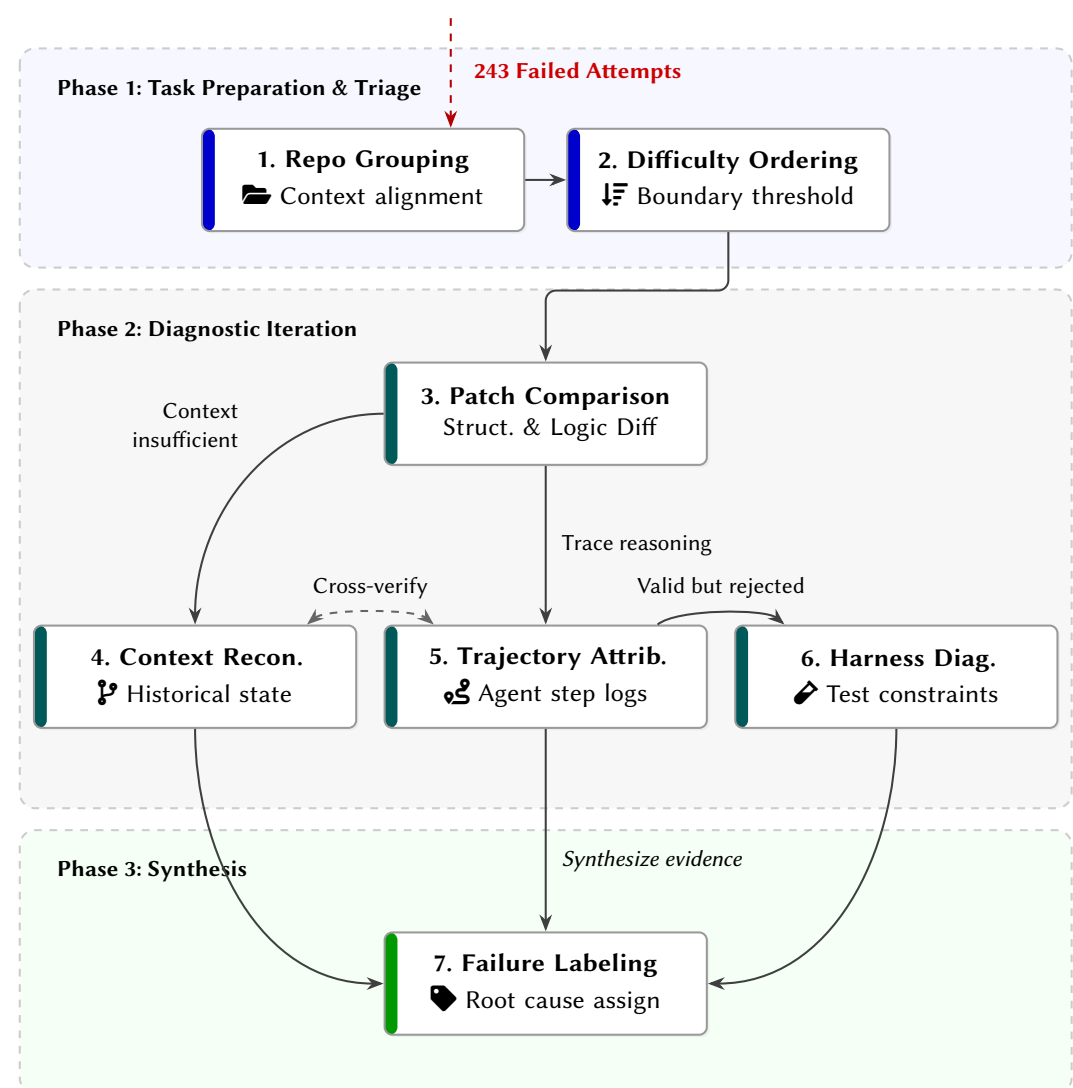

Fig. 2. The failure diagnosis workflow.

**Analysis workflow.** As illustrated in Figure 2, we organize the analysis of the 243 failed issue-resolution attempts into three phases: task preparation and triage, diagnostic iteration, and final synthesis. This structured workflow reduces subjective bias and ensures that all failures are examined under a consistent protocol. To improve annotation reliability, three authors independently inspect the failure cases following the same labeling guidelines. We first conduct a pilot calibration phase on a randomly sampled subset of failures to refine the taxonomy and align annotation criteria. Disagreements are subsequently resolved through group discussion until consensus is reached.

**Phase 1: Task Preparation and Triage.** We first organize tasks to establish a consistent baseline for comparison.

`1. Repository grouping`. Tasks belonging to the same repository are grouped and analyzed together. This reduces context switching and allows reviewers to develop a repository-specific understanding of coding conventions, architectural patterns, and implicit design assumptions.

`2. Difficulty ordering`. Each task contains exactly nine repair attempts (3 models × 3 trials). Within each repository, tasks are analyzed in ascending order of total failure count, starting from tasks with fewer failed attempts and progressing to more difficult ones. This ordering helps analysts gradually build contextual familiarity before examining more complex failures.

**Phase 2: Diagnostic Iteration.** We then perform a detailed diagnosis by comparing generated outputs and tracing the reasoning process.

`3. Patch comparison`. For each task, we systematically compare all failed generated patches against the reference patch. We focus on structural and logic differences, including modified files, insertion locations, content changes, and whether the introduced functionality is behaviorally equivalent to the reference implementation.

`4. Context reconstruction`. When patch comparison alone is insufficient to explain the failure, we reconstruct the repository state using the provided `base_commit` and inspect the exact historical code context. This enables analysis of surrounding control flow, hidden dependencies, and framework-specific constraints that may not be visible from the patch alone.

`5. Trajectory attribution`. We trace the agent's reasoning process step by step using the complete execution logs generated by mini-SWE-agent. By aligning these sequential trajectories with the reconstructed repository context, we identify the earliest point at which reasoning or execution deviates from a correct repair path.

`6. Harness diagnosis`. In cases where the agent successfully passes its self-generated reproduction script but fails the official SWE-bench evaluation, we further inspect the benchmark harness. A patch is considered semantically correct only when multiple reviewers confirm that it satisfies the issue description and aligns with the intent of the reference solution. We then analyze the official `test_patch`, particularly the `FAIL_TO_PASS` and `PASS_TO_PASS` assertions, to determine whether the failure is caused by benchmark strictness, hidden constraints, or unrelated test-side conditions rather than an incorrect repair.

**Phase 3: Synthesis.** Finally, we consolidate all diagnostic evidence and assign failure labels.

`7. Failure labeling`. For each failed attempt, we identify the earliest causally dominant breakdown point in the interaction trajectory and assign a single failure label corresponding to that root cause. Although multiple downstream errors may appear during execution, we use single-label attribution to preserve mutual exclusivity and avoid double-counting. This design allows the taxonomy to focus on the primary source of failure rather than its cascading consequences.

To concretize the manual analysis procedure depicted in Figure 2, we take task `django__django-14771` as a representative example, which requires passing `-X` options to the Django autoreloader. The specific patch content in this case is illustrated in Figure 3. During *Phase 2 (Diagnostic Iteration)*, we execute the workflow as follows. In *Step 3 (Patch comparison)*, we analyze the patch generated by the model (Figure 3a) and the human developer (Figure 3b). Both of them correctly identify the need to extract values from `sys._xoptions` and append them to the subprocess arguments. However, Gemini implements this by appending the flag and value as separate list elements (e.g., `['-X', arg]`), whereas the gold patch concatenates them into a single string (e.g., `f'-X{key}'`). Because the patches are concise and the structural discrepancy is clearly identifiable from the diffs alone, *Step 4 (Context reconstruction)* is unnecessary. Moving to *Step 5 (Trajectory attribution)*, a review of the execution logs reveals that Gemini actually successfully passes its self-generated reproduction script. In cases where the agent

(a) Model Patch (Gemini Round 2)

```
  args = [sys.executable] + ['-W%s' % o for o in sys.warnoptions]
+ if sys.implementation.name == 'cpython':
+ for k, v in sys._xoptions.items():
+ arg = k if v is True else '%s=%s' % (k, v)
+ args.extend(['-X', arg])
  # __spec__ is set when the server was started with the `-m` option,
```

(b) Gold Patch

```
  args = [sys.executable] + ['-W%s' % o for o in sys.warnoptions]
+ if sys.implementation.name == 'cpython':
+ args.extend(
+ f'-X{key}' if value is True else f'-X{key}={value}'
+ for key, value in sys._xoptions.items()
+ )
  # __spec__ is set when the server was started with the `-m` option,
```

(c) Test Patch

```
+ @mock.patch('sys.argv', [__file__, 'runserver'])
+ @mock.patch('sys.warnoptions', [])
+ @mock.patch('sys._xoptions', {'utf8': True, 'a': 'b'})
+ def test_xoptions(self):
+ self.assertEqual(
+ autoreload.get_child_arguments(),
+ [sys.executable, '-Xutf8', '-Xa=b', __file__, 'runserver'],
+ )
```

Fig. 3. Illustration of the diagnostic workflow in django-14771.

successfully passes its self-generated reproduction script but fails the official SWE-bench evaluation, we proceed to *Step 6 (Harness diagnosis)*. Upon inspecting the official test patch (Figure 3c), we discover the underlying issue: the benchmark rigidly asserts exact list equality, expecting the combined string format ['-Xutf8', '-Xa=b']. Although Gemini's separated list syntax is functionally equivalent and valid for Python subprocess execution, the test rejects it.

Finally, in **Phase 3 (Synthesis)** and *Step 7 (Failure labeling)*, we synthesize this collected evidence to assign a definitive root cause. Recognizing that the functionally valid code is rejected due to rigid formatting expectations, we summarize the core issue and assign a descriptive root-cause label, such as "strict format mismatch". This completes the case's diagnostic cycle, serving as a concrete demonstration of how our manual workflow distills raw code files and evaluation traces into precise failure attribution.

## IV. Empirical Results

### A. RQ1: Failure Taxonomy and Distribution

To answer RQ1, we execute the repair process on 100 sampled tasks. Using three LLMs (Claude, Gemini, and GPT) with three independent runs per task, we generate a total of 900 issue-resolution attempts. Following the evaluation process detailed in Section III-E, the benchmark's test suite evaluates these patches. Overall, the models successfully resolve 657 attempts, while 243 (27%) fail to pass the complete evaluation suite. Despite strong overall performance, this non-trivial failure rate highlights the need to better understand the underlying causes of unsuccessful issue-resolution attempts. To delineate their reasoning boundaries and operational bottlenecks, we are diagnosing the root causes of the 243 unresolved instances.

TABLE III
Failure Taxonomy Overview.

| Stage | Symptom | Count |
|---|---|---|
| Problem Understanding | P1 Misinterpretation / Domain Knowledge Lack | 44 |
| | P2 Distracted by Hints / TODOs | 27 |
| Localization | L1 Incomplete Scope / Wrong Layer | 17 |
| Strategy & Logic | S1 Partial Fix / Incomplete Logic | 52 |
| | S2 Incorrect Strategy / Side Effects | 29 |
| | S3 Hardcoding / Bad Practices | 9 |
| Implementation & Execution | I1 Tool Failure / Silent State Hallucination | 4 |
| Validation & Harness Constraints | V1 Specification-Oracle Gap | 10 |
| | V2 Strict Output Format Mismatch | 30 |
| | V3 Side Effects | 17 |
| | V4 Execution Timing Mismatch | 4 |
| **Total** | | **243** |

The results are summarized in Table III. We classified these failures into a taxonomy comprising five stages and eleven fine-grained categories. This taxonomy is empirically derived through an inductive coding process applied to the labels from our manual inspection. Since trajectory logs indicate that agents strictly adhered to the `mini-SWE-agent` workflow (i.e., codebase analysis, reproduction script creation, and iterative editing), our five stages directly map to this agentic progression: Problem Understanding, Localization, Strategy & Logic, Implementation & Execution, and Validation & Harness Constraints. Based on the quantitative distribution in Table III, we highlight the following key findings regarding the failure modes of LLM agents:

1) **Cognitive bottlenecks predominate in the repair process.** The majority of breakdowns occur during cognitive reasoning rather than implementation and execution. The `Strategy & Logic` stage represents the most significant bottleneck (90, 37%). Within this stage, `Partial or incomplete fixes (S1)` constitute the largest category, suggesting that while agents accurately localize faults, they often fail to synthesize comprehensive repair logic. Similarly, `Problem Understanding` accounts for 71 failures (29.2%), typically due to the misinterpretation of issue descriptions or anchoring on misleading hints within the prompt.
2) **Evaluation constraints and harness rigidity impede success rates.** A substantial portion of unresolved cases (40, 16.5%) arise during the `Validation & Harness Constraints` stage (V1&V2). In these cases, the evaluation environment rejects functionally plausible and semantically correct solutions due to rigid assertions or undocumented environmental dependencies, thereby obscuring the models' underlying repair capabilities.
3) **Localization and implementation are no longer primary obstacles.** Operational steps presented minimal friction. `Localization` contributes to only 7% of errors, demonstrating the proficiency of current models in navigating large-scale codebases. Failures in `Implementation & Execution` are nearly negligible (1.6%). While models occasionally encounter fragile code-editing tools, they

**Model Patch (Gemini Round 3: Naive Body Stripping)**

```
1  def finish_response(self):
2      if self.environ["REQUEST_METHOD"] == "HEAD":
3          try:
4              for data in self.result:
5                  if not self.headers_sent:
6                      self.send_headers()
7              if not self.headers_sent:
8                  self.send_headers()
9          finally:
10             self.close()
11     else:
12         super().finish_response()
```

**Gold Patch (Domain-Aware Body & Header Cleanup)**

```
1  def cleanup_headers(self):
2      super().cleanup_headers()
3      if (self.environ["REQUEST_METHOD"] == "HEAD"
4          and "Content-Length" in self.headers):
5          del self.headers["Content-Length"]
6
7  # ... [Connection logic omitted] ...
8  def finish_response(self):
9      if self.environ["REQUEST_METHOD"] == "HEAD":
10         try:
11             deque(self.result, maxlen=0) # Consume iterator.
12             if not self.headers_sent:
13                 self.send_headers()
14         finally:
15             self.close()
16     else:
17         super().finish_response()
```

Fig. 4. Comparison of body and header cleanup handling in django-16502.

consistently bypass these hurdles through custom Python scripts or iterative retries.

The following subsections provide a qualitative analysis of each stage, utilizing illustrative examples to demonstrate typical agentic breakdowns.

*1) **Problem Understanding**:* Failures at this stage stem from an inability to translate natural language descriptions into precise programmatic constraints. As illustrated in Figure 4, models often adopt a surface-level interpretation, focusing on literal wording while overlooking implicit domain rules and architectural hierarchies.

**P1: Misinterpretation/Domain Knowledge Lack.** A common failure pattern arises when tasks require adherence to implicit domain standards, such as the Web Server Gateway Interface specification [44], that are not fully detailed in the prompt. The failure occurs in task `django__django-16502`, where a patch should ensure that `HTTP HEAD` requests suppress the response body while maintaining RFC-compliant headers. As observed in the third-round patch generated by *Gemini*, the model exhibits a surface-level understanding. It identifies the need to suppress the response body but applies a localized, native fix. By overriding `finish_response` to manually exhaust the iterator (lines 4-8), the model successfully strips the body but bypasses the framework's header lifecycle. Consequently, the response still carries a potentially incorrect `Content-Length:0` header. In contrast, the human-written (Gold) patch demonstrates architectural awareness. Instead of merely stripping the body, it intervenes in the `cleanup_headers` hook (lines 1-5) to explicitly remove the `Content-Length` entry before transmission. This dual-layer approach achieves full protocol compliance by efficiently exhausting the iterator and normalizing headers at the correct lifecycle stage. This disparity reveals that while models can mimic functional logic, they struggle with implicit architectural constraints and multi-stage framework interactions.

**Problem Statement (Speculative Hint)**

*"A potential workaround would be to check* `thresholds.max() <= 1` *in which case we should clip* `thresholds` *to not be above 1."*

**Model Patch (Gemini Round 1: Anchoring on Hint)**

```
1  if thresholds.size > 0 and thresholds[0] <= 1:
2      extra_threshold = min(thresholds[0] + 1, 1)
3      if extra_threshold <= thresholds[0]:
4          extra_threshold = thresholds[0] + 1
5  else:
6      extra_threshold = thresholds[0] + 1
7  thresholds = np.r_[extra_threshold, thresholds]
```

**Gold Patch (Mathematically Rigorous)**

```
1  # strictly starting the ROC curve at (0, 0)
2  thresholds = np.r_[np.inf, thresholds]
```

Fig. 5. Impact of misleading textual hints on patch generation in scikit-learn-26194.

**P2: Distracted by Hints/TODOs.** Models exhibit a pronounced susceptibility to misleading contextual cues, such as speculative suggestions in issue descriptions or remaining `TODO` comments in codebases. Rather than performing independent logical verification, LLMs tend to anchor on these informal hints and translate them directly into code implementation. This behavior reveals a critical alignment-induced sycophancy [45], [46], in which the model prioritizes satisfying user-provided intent over safeguarding the software's technical integrity.

A compelling example of this anchoring bias is observed in `scikit-learn__scikit-learn-26194`. As illustrated in Figure 5, the model's reasoning is influenced by the issue reporter's speculative hint, leading to mathematically flawed logic. The task's issue statement contains a textual distractor suggesting that the system should "clip" thresholds if inputs are probability estimates. Rather than evaluating the mathematical validity of this suggestion, Gemini generates a surface-level patch (the red box) that literalizes this hint. The model implements an ad hoc bounding mechanism to ensure the threshold falls within [0,1], exhibiting a fast-thinking pattern that prioritizes textual consistency with the prompt over a functional analysis of the ROC algorithm. The heuristic-driven approach is fundamentally unsound. The objective of prepending an extra threshold in ROC construction is not to constrain values within a valid probability range, but to satisfy an algorithmic invariant, ensuring the curve strictly originates at (0,0). By artificially capping the threshold, the model fails to guarantee this property across diverse score distributions, introducing potentially silent logical regressions.

In contrast, human developers exercise first-principles reasoning by dismissing the misleading hint. As shown in the gold patch (green box), they address the root cause by utilizing positive infinity (`np.inf`). This robustly satisfies the mathematical requirement regardless of the underlying data distribution. This contrast underscores a pivotal gap. While LLMs excel at instruction-following, they lack the conceptual depth and critical skepticism required to navigate misleading context in complex engineering scenarios.

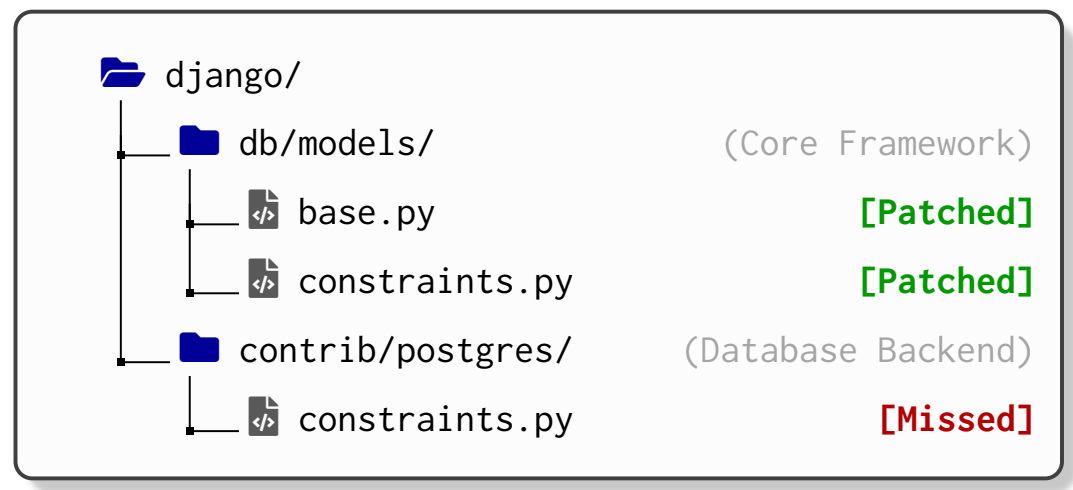

Fig. 6. Localization failure in django-16560.

*2) **Localization**:* Failures at this stage occur when the model correctly understands the problem but fails to identify all code locations required for a complete fix. While pure localization failures are relatively rare (17/243), this does not imply that localization is solved. Instead, models perform well at identifying the primary fault location but struggle when fixes require reasoning across multiple files or architectural layers.

**L1: Incomplete Scope/Wrong Layer.** While LLM agents are proficient at identifying the primary buggy file, they frequently exhibit premature narrowing of the search scope. Consistent with prior findings on the limitations of LLMs in repository-level code comprehension [15], [28], these agents struggle to maintain cross-file and cross-module context. Once a seemingly relevant base class is identified, exploration often terminates early, leading the model to overlook related subclasses or tightly coupled components across architectural layers.

A prevalent manifestation of premature scope narrowing across the evaluated models occurs in task `django__django-16560`. This issue requires introducing a new parameter, `violation_error_code`, across all database constraint implementations. As shown in Figure 6, the evaluated models successfully locate and modify the core abstraction (`BaseConstraint`, defined in *db/models/constraints.py*), together with its shared subclasses within the *django/db/models* module. This suggests that the models can correctly identify the primary locus of the change. However, they fail to recognize that the constraint system extends beyond the core module. In particular, the `ExclusionConstraint` subclass, defined in *contrib/postgres/constraints.py* within the PostgreSQL-specific backend `django/contrib/postgres`, is consistently overlooked. Although it follows the same abstraction contract, it resides outside the model's immediate search scope and is therefore systematically missed. By failing to propagate the modification across architectural boundaries, the resulting patches are incomplete: they function correctly under default database settings but break in PostgreSQL-specific scenarios due to the missing parameter.

*3) **Strategy & Logic**:* Failures at this stage occur when the agent correctly identifies the relevant code region(s) but proposes an incorrect or suboptimal repair strategy. Although the patch location is appropriate, the implemented logic may be incomplete, violate implicit constraints, or introduce unintended side effects. In other words, the resulting patch appears plausible but fails to satisfy the full specification during test execution.

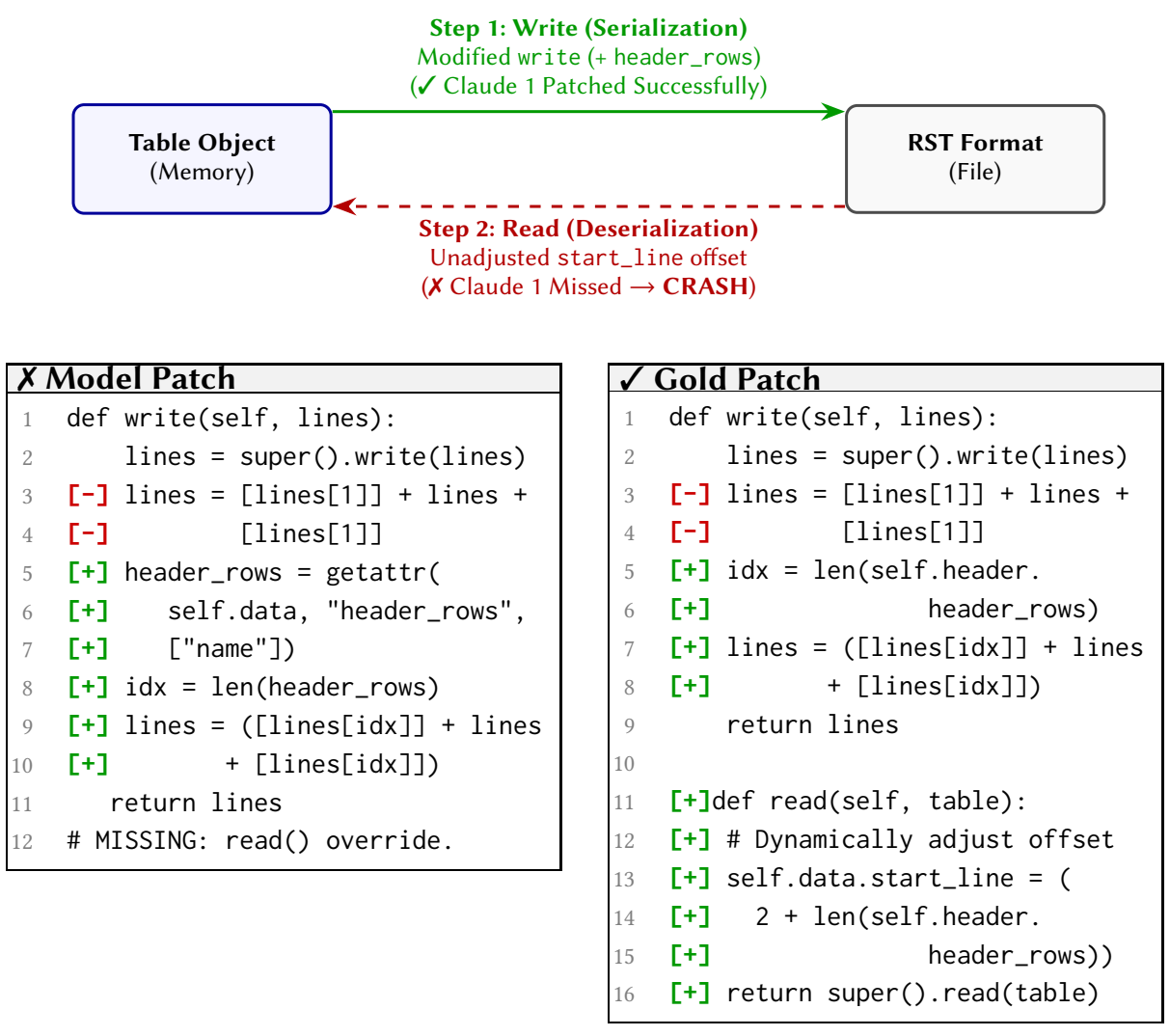

Fig. 7. Bidirectional inconsistency in astropy-14182.

**S1: Partial Fix/Incomplete Logic.** A common failure pattern is that models generate fixes addressing only a subset of the required behaviors. Such patches may resolve the immediate failing test while failing to preserve related behaviors or consistency constraints distributed across the codebase. This reflects a form of test-driven overfitting, in which the model focuses on satisfying observable test failures without fully inferring the broader behavioral expectations implied by the system design [29], [30].

The `astropy__astropy-14182` task serves as a typical demonstration of this incomplete logic. The issue requests support for specifying custom `header_rows` when outputting reStructuredText (RST) tables to resolve a `TypeError`. As illustrated in the conceptual diagram of Figure 7, we analyze the first-round patch generated by Claude. The model successfully modifies the method `write` (the patch on the left), thereby satisfying the explicit requirement for correct serialization (Step 1). However, Claude fails to preserve bidirectional consistency for round-trip operations. In the `astropy` ASCII parser, deserialization relies on a `start_line` offset to separate data rows from headers. Because the model focuses exclusively on the output-side `write` function, the input parser remains unaware of the dynamically inserted header rows. Consequently, when the modified table is read back into memory (Step 2), the parser still begins from the original hardcoded offset, misinterprets the newly inserted string headers as numeric data, and crashes. As shown in the bottom code comparison of Figure 7, the model's patch entirely omits the corresponding deserialization logic. In contrast, the gold

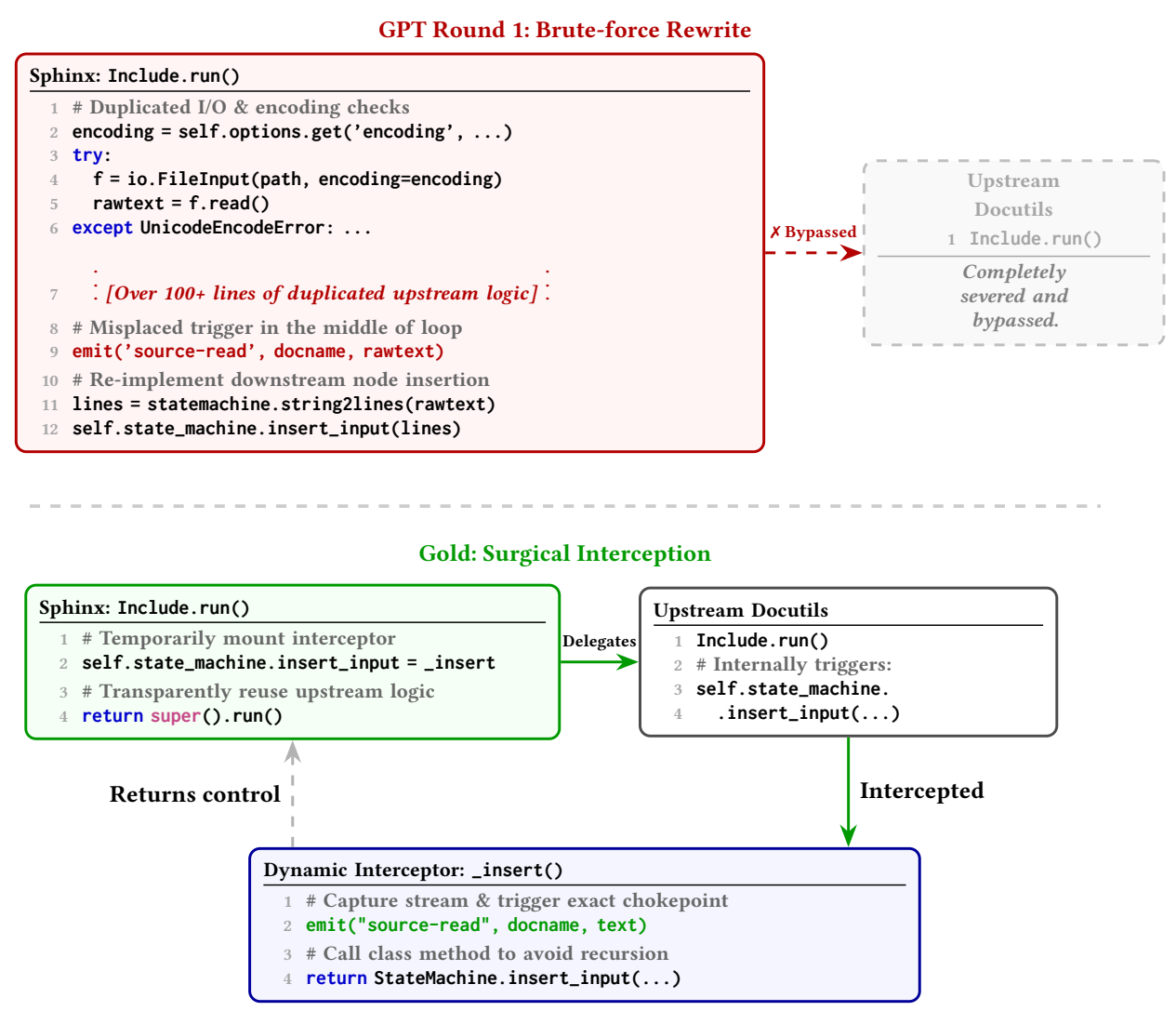

Fig. 8. Comparison of repair strategies in sphinx-11510.

**✗ GPT Round 3 Patch** *(Hardcoding Anti-pattern)*

```
-elif USER_HOME == "~":
-    PYLINT_HOME = ".pylint.d"
 else:
-    PYLINT_HOME = os.path.join(USER_HOME, ".pylint.d")
+    xdg_data_home = os.environ.get("XDG_DATA_HOME")
+    if xdg_data_home:
+        PYLINT_HOME = os.path.join(xdg_data_home, "pylint")
+    elif USER_HOME == "~":
+        PYLINT_HOME = ".pylint.d"
+    else:
+        PYLINT_HOME = os.path.join(USER_HOME, ".local", "share", "pylint")
```

**✓ Gold Patch** *(Platform-Aware Resolution)*

```
-    PYLINT_HOME = os.path.join(USER_HOME, ".pylint.d")
+    PYLINT_HOME = appdirs.user_cache_dir("pylint")
```

Fig. 9. Hardcoding anti-pattern in pylint-4661.

patch explicitly overrides the method `read` to dynamically recompute the correct offset (lines 11-16 in the gold patch). This case exemplifies a classic partial-fix failure: while the model resolves the immediate serialization issue, it overlooks the implicit round-trip consistency contract of the table I/O pipeline.

**S2: Incorrect Strategy/Side Effects.** A prevalent failure pattern in LLM-generated repairs is the adoption of unnecessarily broad modification strategies for localized behavioral changes. This tendency often results in brute-force solutions that prioritize passing the immediate failing tests over long-term maintainability and architectural consistency.

This phenomenon is exemplified in `sphinx-doc__sphinx-11510`. The task requires injecting a `source-read` event into the `include` directive pipeline to synchronize it with the standard source-file workflow. However, as illustrated at the top of Figure 8, the model (GPT Round 1) fails to identify the appropriate extension boundary.

Instead of extending the existing control flow, it performs a monolithic rewrite by transplanting over 100 lines of intricate upstream logic from the *Docutils* library directly into *Sphinx* (red box, line 7). This literal code transplantation not only introduces massive redundancy but also creates a technical debt anchor. By forking a snapshot of upstream logic, the model-generated patch bypasses future security updates and performance optimizations from the parent library, leading to high maintenance overhead and potential logic drift. In contrast, human developers implement a non-intrusive interception (at the bottom of Figure 8). They leverage the existing delegation chain, where `super().run()` eventually calls `insert_input` (green box, line 4). By temporarily wrapping this method with a dynamic interceptor (`_insert`) (green box, line 2), the human patch achieves the desired event trigger at the precise control-flow chokepoint without disrupting the upstream lifecycle. The comparison highlights a critical cognitive gap: while LLMs excel at local code synthesis, they struggle with architectural reasoning across library boundaries. They tend to default to "copy-paste-modify" patterns rather than exploiting the inheritance and interception mechanisms that maintain system decoupling and robust cross-project integration.

**S3: Hardcoding/Bad Practices.** A recurring failure pattern is the models' tendency to bypass generalizable solutions by hardcoding constants or platform-specific assumptions. While such fixes may satisfy immediate test cases, they violate software engineering principles by introducing technical debt and breaking cross-platform compatibility. This hardcoding anti-pattern is presented in `pylint-dev__pylint-4661`. The task requires updating `pylint` to comply with the XDG Base Directory specification when determining its data storage location. As shown in Figure 9, the model (GPT Round 3) attempts a manual reimplementation of path-resolution logic. It explicitly checks the `XDG_DATA_HOME` environment variable and, otherwise, falls back to a hardcoded Linux-specific path `os.path.join(USER_HOME, ".local", "share", "pylint")`. This represents a severe portability bottleneck. By embedding environment-specific hierarchies directly into the patch, the model creates brittle code that ignores OS-specific conventions, leading to deterministic failures in diverse production environments. In contrast, the gold patch leverages ecosystem-aware abstraction. By delegating directory resolution to the `appdirs` utility, the human-written fix transparently handles cross-platform nuances with a single line of code. This abstraction transparently handles platform-specific conventions and reliably returns the correct storage path across operating systems, preserving both portability and maintainability.

It is important to distinguish between two superficially similar failure modes based on where the breakdown occurs in the repair trajectory. **L1 (Incomplete Scope)** refers to a localization-stage failure where the agent identifies the correct fix type but fails to cover the full *edit footprint*, typically reflecting a deficiency in systemic impact analysis. Opera-

**GPT (Round 2)**

```
$ sed -i "s/centers = km.../centers = km...\n
  centers = np.sort(centers)/" sklearn/preprocessing/
_discretization.py
```

*Intention: Multiline replace using \n and hardcoded spaces*

**Bash Environment**

```
<returncode>0</returncode>
<output></output>
```

*Result: Silently fail. Spacing/newline un-matched, file unchanged.*

**GPT (Round 2)**

**THOUGHT:** *"The error persists even after sorting centers... enforce monotonic bin_edges by sorting the final bin_edges array..."*
→ Agent Hallucination & Logic Pivot Triggered

Fig. 10. Agent-environment interaction in scikit-learn-13135 resulting in a silent tool failure and state hallucination.

| Visible Specification (Issue Prompt) | Hidden Oracle (Test Harness) |
|---|---|
| ✓ Explicitly requests fixing `TypeError`<br>✗ **Never mentions** `ValueError` | ✓ Tests `TypeError` recovery<br>✗ **Secretly enforces** `ValueError` |
| **Model Patch (Semantically Aligned):**<br>`except (InvalidOperation, TypeError):`<br>✗ **Rejected by Benchmark** | **Gold Patch (Fitted to Test):**<br>`except (InvalidOperation, TypeError, ValueError):`<br>✓ **Accepted by Benchmark** |

Fig. 11. Specification gap in django-13023.

**Semantic Equivalence vs. Rigid Type Assertion**

*Claude Output:* `str(max_length)` *(String)*
*Test Expects:* `max_length` *(Integer)*

✓ **Functional Execution: Both render identical HTML attributes.**
✗ **Benchmark Oracle: Strict internal type assertion fails.**

Fig. 12. Strict output format mismatch in django-11790.

tionally, we classify a case as L1 when evidence shows additional required edit locations that the agent neither inspected nor modified. In contrast, **S1 (Incomplete Logic)** represents a reasoning-stage failure where the agent targets the correct locations but implements *flawed or partial logic*, failing to satisfy complex invariants or boundary conditions. We classify a failed trial as S1 when the agent edits the correct locations but overlooks critical edge cases or invariants, leaving residual errors. Notably, this divergence reveals a fundamental trade-off in LLM-based repair. L1 represents a failure in the agent's *spatial navigation* of the codebase, which requires enhanced repository-level indexing to resolve. S1 highlights the limits of its *logical depth*, demanding more rigorous formal verification or symbolic reasoning to transcend mere textual alignment with the prompt.

*4) **Implementation & Execution**:* Failures at this stage arise from misalignment between an agent's intended action and its actual execution, typically due to incorrect tool usage or unverified state transitions.

**I1: Tool Failure / Silent State Hallucination.** Models frequently struggle with the syntax of execution tools, generating malformed commands that result in silent failures. In shell environments, certain operations may return a successful exit code (`0`) even when no modification occurs. This leads to state hallucination [47], in which the agent's internal belief diverges from the physical repository's state. The gravity of this failure lies not just in the failed edit but also in the subsequent logic pivot: the agent incorrectly attributes persistent test failures to its reasoning rather than to the tool's execution.

The task `scikit-learn__scikit-learn-13135` provides a representative illustration. The bug involves non-monotonic bin edges in a discretization pipeline that requires sorting the cluster centers. As shown in Figure 10, GPT correctly identifies the root cause and proposes the optimal fix. However, the execution fails during patch application. The agent generates a malformed `sed` command with unescaped newline characters and hardcoded spaces. While the command executes with a `0` status, the pattern mismatch leaves the source code unchanged. Importantly, when subsequent tests continue to fail, the agent experiences a catastrophic misattribution. It assumes its initial hypothesis was flawed and abandons it in favor of a suboptimal fallback, sorting the final `bin_edges`. This case highlights how silent execution errors can corrupt otherwise sound reasoning trajectories.

*5) **Validation & Harness Constraints**:* Failures at this stage expose a fundamental tension between semantic correctness and benchmark evaluation. A non-trivial portion of rejected patches is intent-aligned with the visible requirements, yet fails to satisfy the rigid, underspecified constraints of the evaluation harness. Recent literature highlights that absolute reliance on human-written unit tests as oracles can conflate syntactic matching with semantic correctness [48], [49], potentially leading to an underestimation of an agent's true repair capability [36].

**V1: Specification-Oracle Gap.** This failure mode arises when the evaluation oracle enforces implicit constraints that are absent from the problem description. In such cases, agents perform a locally optimal repair based on the provided context, only to be rejected by a test suite that assumes broader domain knowledge or undocumented edge cases. An instance occurs in task `django__django-13023`. As illustrated in the contrastive analysis in Figure 11, the issue report explicitly characterizes the crash as a `TypeError`. The agent correctly implements an exception handler for the reported symptom. However, the hidden test harness implicitly requires the handler to cover `ValueError` as well. This gap creates a paradox in which agents are penalized for failing to guess hidden constraints. The gold patch's success stems from test-set overfitting and prior developer knowledge, not better reasoning. This underscores the necessity for specification-aware evaluation, ensuring agents are judged only on information available to them during the repair process.

**V2: Strict Output Format Mismatch.** This category represents a particularly inflexible mode of failure, where the evaluation harness enforces strict syntactic or type-level identity rather than functional equivalence. In these scenarios, the model's patch is semantically correct, but it is rejected due to superficial differences in representation. Such cases expose a significant gap between the agent's logical intent and the oracle's rigid internal assertions. An example occurs in task `django__django-11790`. The task involves assigning a maximum length attribute to an HTML widget. As illustrated in Figure 12, the agent successfully implements the fix but casts the value to a string (`str(max_length)`) prior to assignment. From an engineering perspective, it is highly intuitive that HTML attributes are inherently rendered as text. However, the test harness employs a strict internal type assertion, demanding that the value be stored as an integer within the framework's internal dictionary. Although both representations produce identical HTML output, the benchmark marks the task as a failure. The agent is penalized not for a logical error, but for failing to anticipate an implementation detail that has no functional impact on the software's end-to-end behavior.

| Claude 1 Strategy *(Blacklist)* | Gold Strategy *(Whitelist)* |
|---|---|
| **Rule: Cast to float if kind not in 'fc'** | **Rule: Cast to float if kind in 'iu'** |
| **1. Target Bug** *(np.float16, kind 'f')* **Trigger cast ? No → Preserved** *(Fixes local bug)* | **1. Target Bug** *(np.float16, kind 'f')* **Trigger cast ? No → Preserved** *(Fixes local bug)* |
| **2. Unrelated Test** *(bool array, kind 'b')* **Trigger cast ? Yes → Forced Cast** **✗ Breaks global test** | **2. Unrelated Test** *(bool array, kind 'b')* **Trigger cast ? No → Preserved** **✓ Passes global test** |

Fig. 13. Side effect comparison in astropy-8872.

**V3: Side Effects.** A recurring failure pattern in this category is agents' tendency to resolve target issues locally while introducing unintended side effects that destabilize previously passing components. This occurs when an over-generalized repair logic propagates beyond its intended scope, triggering regressions in related modules because the agent fails to account for broader system dependencies and cross-component interactions.

The task `astropy__astropy-8872` exemplifies this phenomenon. In the first attempt by Claude, the agent addresses a precision loss issue where `np.float16` inputs are automatically upcast to `float64`. The model adopts an exclusionary strategy, forcing a cast unless the input is already a floating-point or complex type. While this fix succeeds locally, it lacks the specificity required for library-level code. As illustrated in Figure 13, this broad rule inadvertently captures `boolean arrays`, enforcing them into float representations and breaking established test expectations. In contrast, the gold patch employs an inclusionary strategy. By explicitly restricting casting to a narrow, pre-validated subset of types, the developers isolate the fix to the problematic cases while safely preserving the behavior of unrelated data types. This case underscores a critical limitation that agents often optimize for local textual compliance but fail to perform the impact analysis necessary to maintain global system invariants.

**Execution Timing Comparison**

| Execution Phase | Model Patch | Gold Patch |
|---|---|---|
| **1. Initialization** | Lazy eval via `@property` Creates a proxy object. | Immediate eval in `__init__` Backups original ref. |
| **2. Reflection** | ✗ Crash: Accesses proxy. | ✓ Pass: Accesses instance. |
| **3. Serialization** | *Unreachable after crash.* | ✓ Pass: Uses backup ref. |

Fig. 14. Execution timing mismatch in `django-13343`.

**V4: Execution Timing Mismatch.** This category captures failures where a syntactically valid patch executes its logic at the wrong point in the program's runtime lifecycle. Models frequently defer execution (e.g., via lazy evaluation) rather than eagerly evaluating expressions during initialization. While functionally plausible, this timing mismatch violates the framework's implicit temporal contracts, causing downstream components that expect pre-initialized concrete objects to fail.

An illustrative instance occurs in `django__django-13343`. The issue requires preserving a callable `storage` parameter without evaluating it during migration serialization, while still exposing the evaluated instance during normal runtime. As shown in Figure 14, the model employs a lazy evaluation strategy (Claude Round 2). It converts the `storage` attribute into a dynamically computed `@property`, effectively deferring execution until the attribute is explicitly accessed. However, the framework's internal reflection engine expects a fully initialized concrete object immediately after instantiation. When the framework inspects the field and encounters a delayed property descriptor instead of the actual object, it triggers a `Type Mismatch` crash. In contrast, the gold patch adopts an eager execution strategy coupled with state backup, explicitly satisfying both the immediate initialization requirement and the future serialization contract.

> **Summary of RQ1:** The failure distribution is heavily skewed toward cognitive reasoning rather than mechanical operations, suggesting that repository navigation is no longer the primary hurdle. While LLMs align with human cognitive biases through shallow, partial fixes, they diverge significantly by introducing novel failure modes, such as silent state hallucinations and susceptibility to benchmark rigidity. These results indicate that AI introduces unique bottlenecks linked to information processing and evaluation constraints, shifting the repair frontier from tool-use proficiency to semantic depth.

### *B. RQ2: Characterization of Model Behavioral Patterns*

To understand how different large language models perform in automated issue resolution, we compare three representative models (Claude 4.5 Sonnet, Gemini 3 Pro, and GPT 5) across three dimensions: failure distribution, execution stability, and trajectory efficiency. Our analysis yields four key findings that characterize the processual integrity of these agents.

**Finding 1: Models exhibit distinct reasoning bottlenecks during repair.** As shown in Figure 15, the models differ not only in total failure counts (Claude: 70, Gemini: 82, GPT: 91) but also in their error distributions across the repair lifecycle. Claude's failures are relatively balanced between Problem Understanding (32.9%) and Strategy & Logic (31.4%), suggesting its limitations arise equally from task interpretation and solution formulation. In contrast, Gemini and GPT exhibit a clear concentration of failures in Strategy & Logic (41.5% and 37.4%, respectively), indicating that synthesizing complex repair logic rather than understanding the issue is their primary bottleneck. Within the Strategy&Logic stage, the models show distinct behavioral patterns. GPT produces the highest number of Partial Fixes (Category S1, 21 cases), suggesting it often addresses the immediate symptom but fails to generalize to edge cases. Gemini, on the other hand, records the most Incorrect Strategies (Category S2, 13 cases), indicating a tendency to select overly complex or invasive modification strategies that introduce new regressions.

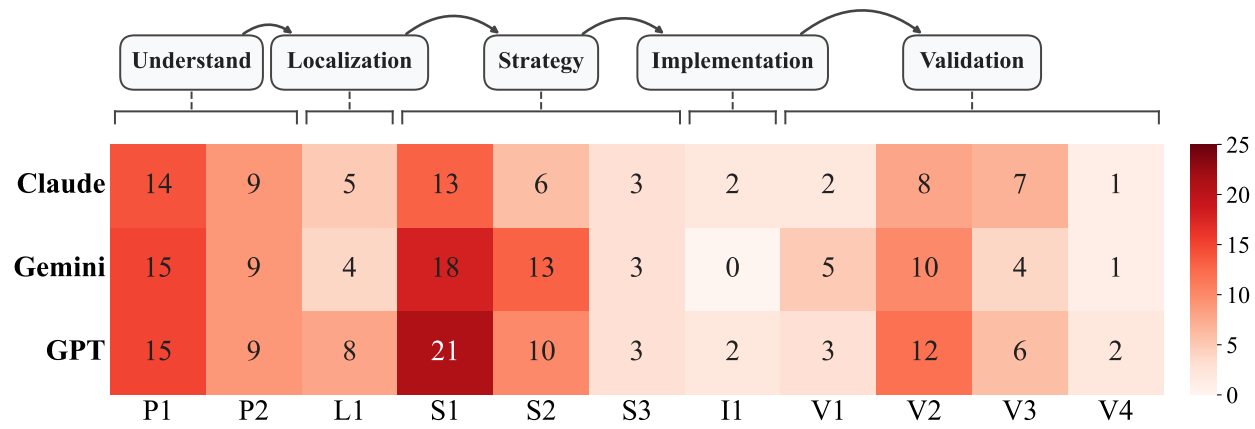

Fig. 15. Heatmap of failure mode distributions.

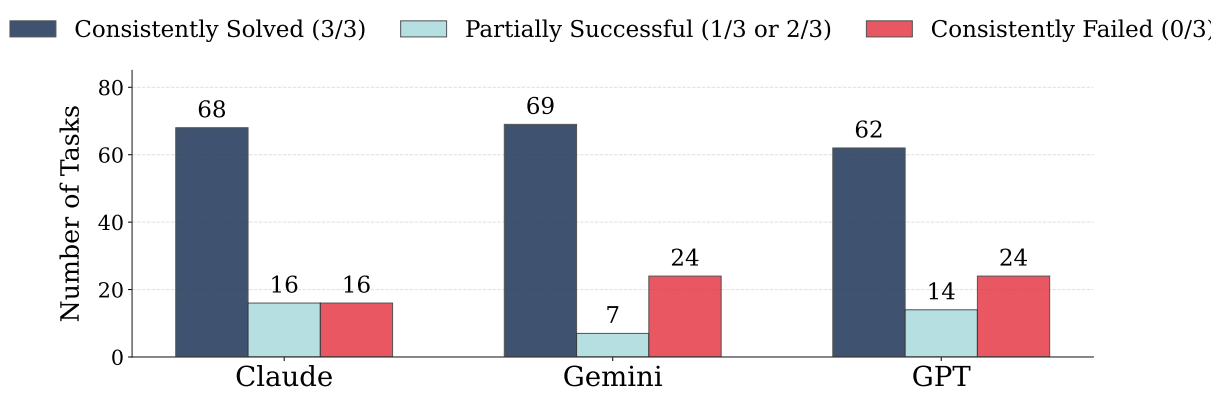

Fig. 16. Execution consistency Comparison.

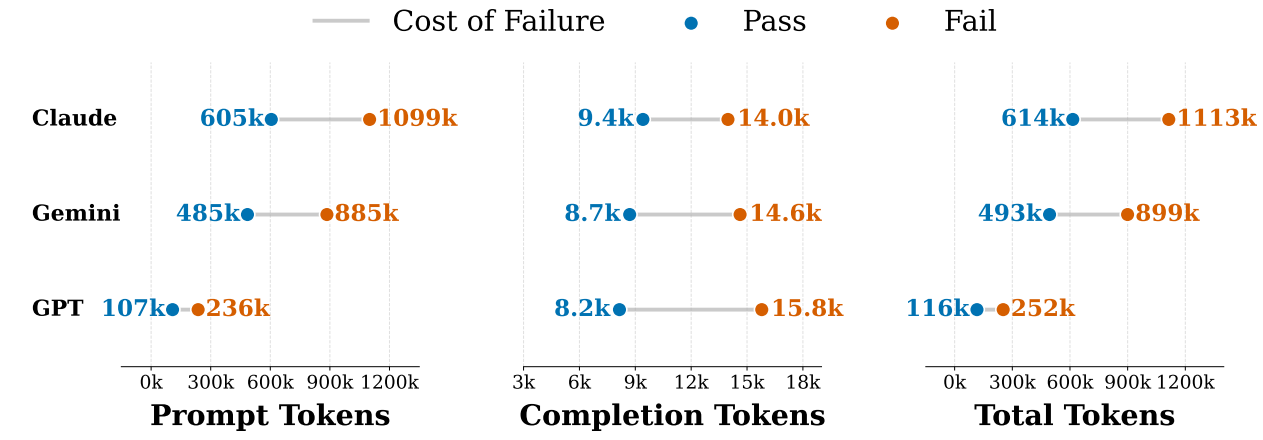

Fig. 17. Trajectory overhead Comparison.

**Finding 2: All models share common vulnerabilities to misleading context and rigid evaluation constraints.** Despite architectural differences, all models encounter similar blind spots. In the Validation & Harness Constraints stage, each model exhibits a comparable failure rate (approximately 25%), suggesting these errors are driven by the inherent rigidity of the test environment rather than model-specific weaknesses. Additionally, we observe a striking uniformity in susceptibility to misleading contextual signals. Each model fails exactly 9 times due to distraction from user hints or legacy `TODO` comments (Category P2), and 3 times due to hardcoded solutions (Category S3). This suggests that alignment to surface-level cues remains a shared fundamental limitation across frontier models, regardless of their reasoning depth.

**Finding 3: Gemini achieves the highest execution stability, whereas Claude and GPT exhibit greater stochastic variance.** Given the stochastic nature of agentic workflows, we evaluate execution stability using three independent trials per task. We categorize outcomes by the number of successful runs: consistently solved (3/3 runs), partially successful (1/3 or 2/3 runs), or consistently failed (0/3 runs). As illustrated in Figure 16, Gemini demonstrates the most stable behavior, consistently solving 69 tasks with only 7 partially successful cases. This high reliability suggests that Gemini is less sensitive to stochastic variations, making it a more predictable candidate for automated CI/CD integration. Conversely, Claude and GPT exhibit higher variability (16 and 14 partially successful cases, respectively). This variance indicates that single-run evaluations significantly underestimate their peak capabilities, as successful patches often emerge only across multiple attempts.

**Finding 4: Models suffer from reasoning inertia and severe inefficiency during failed trajectories.** We analyze efficiency using Prompt and Completion Tokens as proxies for context ingestion and reasoning effort. As shown in Figure 17, GPT is the most targeted model, resolving tasks with significantly fewer prompt tokens (≈107k) compared to Claude (≈605k) and Gemini (≈485k), which rely on massive context windows. However, a consistent inefficiency emerges during failures. Across all models, prompt token usage increases sharply during failed runs, suggesting a lack of effective early stopping mechanisms. Instead of recognizing an unsolvable path, models fall into a state of reasoning inertia (repetitive, unfocused debugging loops). Notably, GPT's verbosity doubles during failures (from ≈8.2k to ≈15.8k completion tokens), exceeding both Claude and Gemini. This suggests that while GPT attempts to reason its way out through extended self-correction, these additional steps yield diminishing returns and rarely resolve the underlying logical flaws.

> **Summary of RQ2:** Models exhibit distinct reasoning bottlenecks but share common vulnerabilities to misleading hints and rigid test harnesses. Gemini provides the highest execution stability, offering better predictability for deployment. Crucially, all models suffer from severe inefficiency during failed attempts, where they fall into unproductive reasoning loops that consume massive computational resources without correcting the underlying errors.

### C. *RQ3: Task Features and Actionable Insights*

Building on the failure taxonomy, we investigate the intrinsic task-level characteristics that trigger these failure boundaries and derive mitigation strategies. We categorize our analysis into two cohorts: (1) Reasoning and Strategic Failures, where we analyze tasks that consistently failed across all trials to identify hard reasoning limits, and (2) Evaluation Constraints, where we examine the entire subset to characterize the strictness gap between models and benchmarks. By identifying these recurring patterns, we aim to provide actionable insights for developers, enabling them to recognize similar risk factors in real-world applications and proactively apply the mitigation strategies suggested by our taxonomy.

*1) P1: Implicit Rules and Knowledge Boundaries:* `Task Features:` These tasks are characterized by information vacuums. They depend on domain-specific rules (e.g., cryptographic protocols, complex formatting specifications) absent from both the codebase and the issue description. `Insight:` Textual hints are insufficient. Models cannot reason their way through missing specifications. `Actionable Strategy:` Autonomous agents should be equipped with dynamic documentation retrieval (e.g., RAG over official APIs) rather than relying solely on pre-trained internal representations.

*2) P2: Textual Distractors and Alignment Bias:* `Task Features:` These tasks contain surface-level seductions, which refer to highly visible but incorrect solution hints within bug reports or legacy TODO comments. `Insight:` Models exhibit a strong alignment bias, prioritizing explicit human suggestions over implicit system logic. `Actionable Strategy:` Implement a logic-first verification step where agents are explicitly prompted to treat user-provided hints as untrusted hypotheses to be validated against core system invariants.

*3) L1: Structural Dispersal and Boundary Traps:* `Task Features:` These tasks involve scattered impact zones where logic is spread across disconnected modules (e.g., base components and isolated backends). `Insight:` A narrow problem description acts as a boundary trap, causing the agent to focus on a local visible fix while ignoring identical defects in distant directories. `Actionable Strategy:` Enhance localization agents with global cross-reference analysis and similar code search patterns to uncover hidden implementation replicas.

*4) S1: Functional Symmetry and State Propagation:* `Task Features:` Defined by implicit contracts, these tasks involve paired operations where a change in one side demands a reciprocal update. Failures often manifest as isolated crashes far from the root cause. `Insight:` Agents tend to apply superficial patches at the crash site rather than maintaining the integrity of the entire data pipeline. `Actionable Strategy:` Introduce data-flow integrity checking that requires agents to trace the full lifecycle of a modified object before finalizing a patch.

*5) S2: Foundational Coupling and Ripple Effects:* `Task Features:` These tasks touch the architectural bedrock, which refers to low-level parsers or query builders with massive downstream dependencies. `Insight:` The risk of regression is extreme because the core changes affect hundreds of modules. `Actionable Strategy:` For core-component modifications, agents should prioritize exploratory regression testing over local validation, explicitly scanning for side effects in distant test suites.

*6) V1&V2: The Oracle-Specification Gap:* `Task Features:` A profound mismatch between flexible user expectations and inflexible test oracles. Tests often demand absolute equality of internal types or string formats that are never specified in the task. `Insight:` These are artificial failures, and the repair is functionally correct but violates a hidden, arbitrary design choice of the original developer. `Actionable Strategy:` Evaluation frameworks should adopt semantic oracles or provide environment-aware hinting to expose hidden constraints during the reasoning phase.

**Validating Targeted Interventions Based on Task Features.** To further examine whether certain failure modes arise from artificial bottlenecks rather than fundamental reasoning limitations, we conduct targeted prompt interventions on the subset of tasks that are consistently failed in our RQ3 analysis. We focus on three categories (V1, V2, and P1) because our qualitative findings suggest that these failures are primarily due to either contextual information gaps or missing domain-specific knowledge. Our goal is not to propose a practical prompting strategy, but rather to use controlled interventions as a diagnostic tool for causal analysis. By explicitly exposing hidden constraints to the agents, we test whether the models can generate correct patches once the missing information is provided.

For V1 (Specification-Oracle Gap) and V2 (Strict Output Format Mismatch), the interventions prove effective. Across the filtered task subset, we observe substantial improvements in repair success after introducing explicit task constraints, indicating that these failures are largely prompt-sensitive rather than capability-bounded. Below, we present two illustrative examples. In django-13023 (V1), models consistently fail because the hidden test suite requires handling `ValueError`, while the issue description explicitly mentions only fixing a `TypeError` [37], [50]. To simulate a setting in which this hidden constraint is made visible, we append an additional instruction warning the model that issue descriptions may be incomplete and that hidden evaluations often test broader behavioral boundaries than those explicitly described. With this targeted guidance, all evaluated models proactively extend their exception handling to cover both `TypeError` and `ValueError`, successfully passing the previously failed evaluation. This result suggests that the failure does not stem from an inability to implement the correct logic, but rather from incomplete task specification.

A similar pattern is observed in django-11790 (V2) [37], [51]. The benchmark for this task strictly enforces an integer type for an HTML attribute, penalizing models that generate functionally equivalent string assignments. We therefore append a task-specific warning emphasizing that strict equality assertions require preserving native data types and that framework helper functions may implicitly cast integers into strings. This explicit boundary condition effectively recalibrates model behavior. Gemini, for example, avoids helper functions that trigger implicit string conversion and instead preserves the required integer type, improving from zero successful attempts to a perfect pass rate on this task. Together, these cases indicate that once hidden constraints and strict formatting requirements are explicitly surfaced, resolving V1 and V2 failures falls well within the capabilities of current models.

**Summary of RQ3:** Our analysis reveals that persistent failures are not random but are structurally tied to features such as architectural dispersal, functional symmetry, and undocumented domain rules. While alignment gaps can be effectively mitigated through targeted context exposure, hard knowledge boundaries remain resilient to prompt-based interventions. This underscores a critical evolution for future repair agents: moving beyond pure reasoning toward active knowledge acquisition and logic-driven verification to bridge the expertise gap between AI and human experts.

## V. Discussion

### A. Implications for Future Research and Practice

While isolated prompt engineering can mitigate specific bottlenecks, achieving robust autonomous repair requires a fundamental transition in methodology. Based on our empirical analysis of 243 failure cases, we synthesize five implications for the academic and industrial software engineering communities.

**Implication 1: Transitioning from code retrieval to systematic planning.** Current coding frameworks, such as AutoCodeRover [12] and Agentless [13], heavily prioritize optimizing fault localization. However, our taxonomy reveals that pure localization errors (L1) account for less than 8% of failures in frontier models. The primary bottleneck lies in Strategy&Logic (S1, S2), which accounts for up to 40% of failures in models like Gemini. Our findings suggest that even when a model correctly identifies the buggy file, it often lacks the capacity to synthesize a coherent fix, frequently producing invasive rewrites (e.g., `sphinx-11510` [52]) instead of minimal, human-like solutions. Future architectures should move beyond retrieval-centric designs and explicitly incorporate structured reasoning or self-reflection to validate logic before attempting code modification.

**Implication 2: Mitigating user compliance bias through critical skepticism.** We observe a recurring failure pattern where models anchor on speculative user hints or misleading legacy comments (P2). This alignment-induced sycophancy causes models to prioritize prompt adherence over logical correctness, leading to the implementation of flawed workarounds. To enhance reliability, agent designs ought to enforce critical skepticism. Implementing a specification refinement stage or multi-agent debate [53] could help verify problem descriptions against the actual control flow, ensuring that models defend against, rather than propagate, user-introduced errors.

**Implication 3: Overcoming the specification-oracle gap in code evaluation.** While traditional studies focus on patch overfitting [54], [55], our data on Validation&Harness Constraints (V1, V2) highlights the opposite challenge. Rigid test oracles often reject functionally correct but syntactically divergent patches. This specification-oracle gap suggests that relying solely on human-written unit tests may unfairly penalize valid repairs. The community needs to explore execution-based semantic evaluators or utilize LLMs as judges [49] to analyze trace semantics, moving toward a more nuanced assessment of patch correctness that transcends simple string matching.

**Implication 4: Addressing execution randomness for rigorous benchmarking.** Our repeated trials demonstrate that autonomous repair is highly sensitive to the random nature of auto-regressive generation. While Pass@1 [1], [31] is the common metric due to cost constraints, our results show that it often conflates a model's intrinsic capability with favorable search trajectories, particularly for models like Claude. It is imperative for future empirical evaluations to report Pass@k and stability metrics. This approach will provide a more accurate representation of a model's reliability and its true performance ceiling in real-world repository repair.

**Implication 5: Bridging domain knowledge gaps via active documentation retrieval.** Our analysis indicates that while prompt engineering can resolve formatting issues, it is inadequate for addressing deep domain knowledge deficits (P1). Abstract instructions cannot replace the concrete and framework-specific knowledge required for complex repairs, highlighting a hard boundary in a model's parametric memory. To bridge this gap, future systems should integrate retrieval-augmented generation tailored for external developer documentation and API references. Grounding agents in official system specifications rather than relying on pre-trained memory is essential for resolving architectural-level defects in large-scale software repositories.

### B. Threats to Validity

**Internal Validity.** The primary concern involves the potential for subjective bias during the root-cause analysis of agent failures. To mitigate this threat, we implement a labeling protocol where three authors independently scrutinize each repair trajectory, cross-referencing execution logs, code diffs, and reference patches. Any labeling discrepancies are resolved through consensus-based group discussions guided by a standardized, evidence-based checklist to ensure taxonomic consistency. Furthermore, while we impose a 250-step interaction limit and a $5.00 cost cap per task, our trajectory logs indicate that all attempts concluded with an autonomous submission by the model rather than a forced termination. This suggests that resource constraints did not fundamentally distort the observed failure patterns.

**External Validity.** The generalizability of our findings is naturally bounded by the choice of dataset and programming language. Our study relies on the SWE-bench Verified benchmark, which consists of Python projects. Due to Python's interpreted nature, we may have observed fewer mechanical implementation or environment-related errors than in compiled languages like Java or C++, where build system complexities are more pronounced. Additionally, while our sample of 100 tasks is representative of the benchmark's repository diversity, it may not fully capture the specific failure modes associated with extremely long-horizon debugging or large-scale architectural migrations.

**Construct Validity.** To account for potential test flakiness, we execute the official evaluation harness three times for each generated patch. The results remain deterministic across all runs, confirming that performance variance stems from the random nature of LLM trajectory generation rather than environment instability. We also acknowledge the limitations of relying on unit tests as the sole oracle. This approach may fail to credit semantically correct patches that diverge from strict test assertions, a phenomenon that likely contributed to the specification-oracle gap identified in our analysis. Furthermore, recent studies suggest that evaluation results on SWE-bench may be inflated due to limitations in test suites. As a result, passing all tests does not necessarily guarantee full functional correctness. Since our analysis focuses on failed repair attempts, incorrect patches that pass the tests may be excluded, potentially leading to an incomplete characterization of failure modes. We partially account for this limitation by incorporating the specification-oracle gap into our failure taxonomy.

## VI. Conclusion

This study provides a systematic deconstruction of the resolution gap between frontier Large Language Models and human developers in repository-level software issue resolution. Through a rigorous manual root-cause analysis of 243 failure trajectories, we demonstrate that the primary bottleneck in autonomous resolution has shifted from mechanical tool execution and fault localization toward high-order cognitive reasoning. While frontier models exhibit remarkable proficiency in identifying faulty code regions, they frequently falter during strategy formulation and logic synthesis. These failures are often exacerbated by an alignment bias, where models anchor on misleading textual hints at the expense of system-level invariants. Furthermore, we uncover a critical specification-oracle gap, revealing that 16.5% of failures stem from rigid evaluation harnesses that penalize functionally correct patches. This suggests that current benchmarks may systematically underestimate model capabilities, highlighting an urgent need for semantic-aware evaluation frameworks. To achieve human-level proficiency, future research should pivot from retrieval-centric architectures toward logic-driven agents that prioritize systematic planning, critical skepticism, and active knowledge elicitation.

## Acknowledgements

This work was partially supported by the National Natural Science Foundation of China (62502343, 62232003). This publication has emanated from research conducted with the financial support of Taighde Éireann–Research Ireland under Grant number 13/RC/2094_2. Lionel Briand is also supported by the Natural Sciences and Engineering Research Council of Canada.